# Multi-level Phenotypic Models of Cardiovascular Disease and Obstructive Sleep Apnea Comorbidities: A Longitudinal Wisconsin Sleep Cohort Study

Duy Nguyen[1#], Ca Hoang[2#], Phat K. Huynh[1,2]*, Tien Truong[2], Dang Nguyen[3], Abhay Sharma[4], and Trung Q. Le[2,5]*

[1] *Department of Industrial and Systems Engineering, North Carolina A&T State University, Greensboro, NC 27411, USA*

[2] *Department of Industrial and Management Systems Engineering, University of South Florida, Tampa, FL 33620, USA*

[3] *Massachusetts General Hospital, Corrigan Minehan Heart Center, Harvard Medical School, Boston, MA 02114, USA*

[4] *Department of Otolaryngology–Head and Neck Surgery, University of South Florida, Tampa, FL 33612, USA*

[5] *Department of Medical Engineering, University of South Florida, Tampa, FL 33620, USA*

[#] *Co-first authors*

*\* Co-Corresponding authors: Phat K. Huynh, email: huynh105@usf.edu, Trung Q. Le, email: tqle@usf.edu*

## *Abstract*

Cardiovascular diseases (CVDs) are prevalent among obstructive sleep apnea (OSA) patients, presenting significant challenges in predictive modeling due to the complex interplay of comorbidities. Current methodologies predominantly lack the dynamic and longitudinal perspective necessary to accurately predict CVD progression in the presence of OSA. This study addresses these limitations by proposing a novel multi-level phenotypic model that analyzes the progression and interaction of these comorbidities over time. Our study utilizes a longitudinal cohort from the Wisconsin Sleep Cohort, consisting of 1,123 participants, tracked over several decades. The methodology consists of three advanced steps to capture the nuanced relationship between these comorbid conditions: (1) Performing extensive feature importance analysis using tree-based models, which highlight the predominant role of variables in predicting CVD outcomes. (2) Developing a logistic mixed-effects model (LGMM) to identify longitudinal transitions and their significant factors, enabling detailed tracking of individual trajectories; (3) Utilizing t-distributed Stochastic Neighbor Embedding (t-SNE) combined with Gaussian Mixture Models (GMM) to classify patient data into distinct phenotypic clusters. In the feature importance analysis, clinical indicators such as total cholesterol, low-density lipoprotein (LDL), and diabetes emerged as the top predictors, highlighting their significant roles in CVD onset and progression. Our predictive models exhibited a high diagnostic accuracy with the LGMM achieving an aggregate accuracy of 0.9556. This model effectively differentiated between patients with varying risk profiles, leading to precise predictions of CVD events. The phenotypic analysis identified two distinct clusters, each representing different risk levels and pathways of disease progression. Specifically, one phenotype was associated with a significant increased risk of major adverse cardiovascular events (MACEs) compared to the baseline, confirming the critical predictive value of nocturnal hypoxia and sympathetic nervous system activation derived from sleep monitoring data. The transition and trajectory analysis of patients using t-SNE and GMM clustering provided concrete evidence of the progression patterns within the cohort. Patients in Cluster 1 showed slower progression to severe CVD states compared to those in Cluster 2, who exhibited rapid onset and deterioration. Conclusively, our study provides a deep understanding of the dynamic interactions between CVD and OSA, offering robust tools for predicting CVD onset and informing personalized treatment strategies.

## I. INTRODUCTION

Obstructive sleep apnea (OSA) —a condition characterized by episodes of upper airway obstruction during sleep—is a prevalent yet commonly undiagnosed and undertreated comorbidity within the population of patients with cardiovascular diseases (CVDs) [1-4]. Alarmingly, 40% to 80% of patients with CVDs, encompassing major adverse cardiovascular events (MACEs) categories [5]: acute coronary syndrome/ischemic heart disease, chronic heart failure, cerebrovascular accidents, and arrhythmias, suffer from OSA. This comorbidity escalates overall morbidity and increases the risk of premature all-cause mortality [3, 6-9]. Moreover, the coexistence of OSA and CVDs poses a considerable societal and economic burden, with healthcare-related expenses attributed to OSA reaching $150 billion and an extra $30 billion incurred when considering CVD-related comorbidities [10]. In particular, the interrelation between OSA and various forms of cardiovascular illness such as hypertension, coronary artery disease, and heart failure has been well-documented [11-17]. The mechanisms proposed to underlie this association include metabolic dysregulation [11], oxidative stress from intermittent hypoxia [14, 18], and altered autonomic nervous system activity [14, 19], which collectively contribute to cardiovascular pathology in OSA patients. However, the complexity of these interactions, compounded by a scarcity of longitudinal data, makes it challenging to untangle the temporal and causal relationships between these comorbid conditions [9, 20, 21]. This complexity is further magnified by the variability in individual patient responses and the multitude of confounding factors that obscure the direct impacts of these diseases on each other [11, 22].

Numerous studies have rigorously explored the connections between OSA and CVDs, consistently highlighting a robust association that manifests through a range of physiological disruptions predisposing patients to cardiac ailments. Particularly, OSA's hallmark feature of intermittent hypoxia has been linked to significant cardiovascular consequences including systemic hypertension, atrial fibrillation, and heart failure, emphasizing the severity of its impact [14, 18, 23-25]. Underlying these clinical manifestations are complex physiological mechanisms involving not only the intermittent hypoxia but also metabolic dysregulation and the imbalance between sympathetic and parasympathetic nervous system activities [26]. Metabolic dysregulation in OSA patients often manifests as disrupted glucose metabolism and lipid profiles, which are well-known risk factors for CVDs [27, 28]. Additionally, the recurrent oxygen desaturation followed by reoxygenation characteristic of OSA induces oxidative stress, which further exacerbates endothelial dysfunction, thus promoting atherosclerosis [29, 30]. Sympathetic nervous system overactivity, coupled with reduced parasympathetic tone, also contributes to the cardiovascular burden by increasing heart rate and blood pressure, thereby imposing greater cardiac workload and risk of myocardial injury [31, 32]. Despite extensive documentation of these relationships, much of the existing literature has relied on cross-sectional or observational studies, which can only infer associations rather than causal mechanisms. This has left significant gaps in our understanding of how these physiological disruptions progress over time to exacerbate or possibly initiate CVD pathology in patients diagnosed with OSA [33].

Despite the observed OSA-CVD associations, there remains a notable deficiency in longitudinal research that probes the evolution and interaction of these conditions over extended periods. Current literature predominantly comprises cross-sectional, one-size-fits-all modeling approaches that fail to capture endophenotype-specific dynamic physiological changes associated with OSA, such as intermittent hypoxia-induced oxidative stress and fluctuations in autonomic nervous system activity, which are critical in the development and progression of CVDs [6, 34-36]. This gap significantly hampers our capacity to predict critical transitions in disease states, thereby limiting the effectiveness of timely interventions aimed at preventing severe cardiovascular incidents. Furthermore, the existing predictive longitudinal models for OSA-CVD comorbidity often ignore the complex, multifactorial nature of OSA-CVD interactions [3, 37-40]. These models typically simplify the influence of genetic predispositions, environmental factors, and lifestyle choices, which are integral to the onset and progression of these diseases. As a result, such models do not adequately capture the heterogeneity of patient experiences and responses to treatment, which are crucial for personalizing medical care. Moreover, the methodology commonly employed in studying these diseases seldom integrates advanced statistical techniques that could handle the dynamic complexity of multilevel biological and clinical data. The utilization of such methodologies is imperative not only for understanding how treatment regimens affect disease progression over time but also for identifying potential early indicators of disease exacerbation. There is a pressing need for endophenotype-specific predictive models that can incorporate time-varying covariates, account for inter-individual variability, and provide a robust model for simulating disease progression scenarios under various intervention strategies.

To overcome the limitations, our paper proposes a novel integration of logistic mixed-effects models (LGMM) with advanced machine learning tools such as t-SNE and Gaussian Mixture Models (GMM). This hybrid approach enables the dynamic analysis of CVD progression in OSA patients over time, addressing a crucial gap in existing research that largely depends on static, cross-sectional data. By employing these sophisticated analytical techniques, our study provides a more profound understanding of the temporal dynamics and phenotypes within OSA-CVD comorbid conditions. In addition, our research methodology identifies and validates new predictive biomarkers and phenotypic patterns associated with CVDs in OSA patients. These findings are crucial for the early detection of CVDs and enable targeted interventions, significantly contributing to the field by enhancing diagnostic precision and treatment efficacy. By identifying specific phenotypes that correlate with effective treatment responses, this study paves the way for the development of personalized treatment strategies.

## II. METHODOLOGY

Our methodology for investigating the longitudinal associations of OSA and CVD within the Wisconsin Sleep Cohort Study (WSCS) [41, 42] comprises three distinct steps, as illustrated in **Figure 1**. Initially, we preprocess and engineer features from the WSCS dataset. This stage involves categorical encoding, KNN-based imputation, outlier detection, and SMOTE for handling class imbalances by up-sampling the minority group by 100%. We then select the top 20 features using a tree-based feature ranking method. The second step employs LGMM to accommodate both fixed and random effects across patient timelines, facilitating the detailed examination of CVD progression. Next, we employ the t-SNE technique for dimensionality reduction and the GMM for OSA-CVD phenotypic clustering. Lastly, our model is validated using different evaluation metrics and expert clinical knowledge.

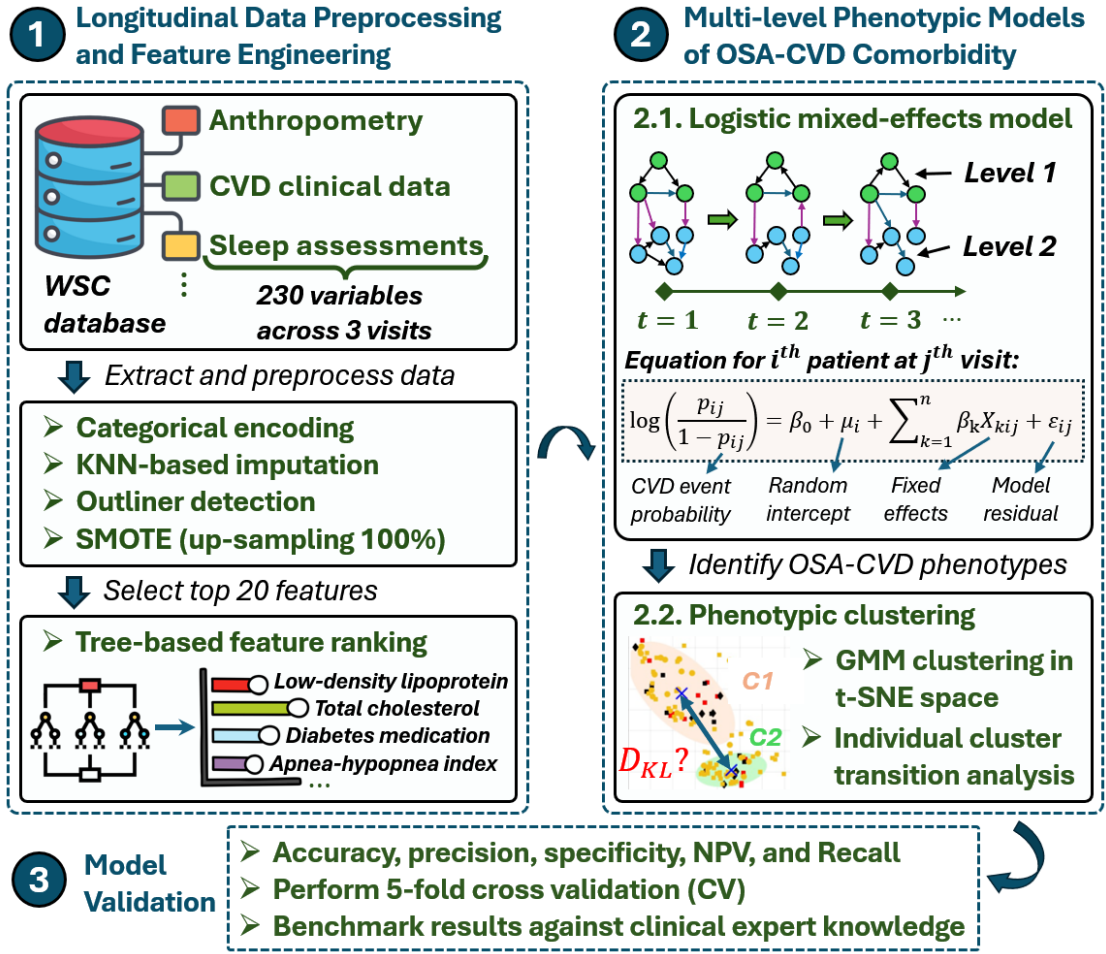

**Figure 1**. Overview of the 3-step methodological framework for modeling OSA-CVD comorbidity utilizing WSC database. Step 1 focuses on data preprocessing and feature engineering, step 2 details the logistic mixed-effects model and phenotypic GMM clustering, and step 3 outlines model validation approach.

## A. Wisconsin Sleep Cohort Study Description and Our Longitudinal Study Design

The study employs a longitudinal design utilizing data from the WSCS, which was initiated in 1988 to investigate the epidemiology and long-term health consequences of sleep disorders, with a particular focus on OSA. The WSC, renowned for its depth and longitudinal nature, includes comprehensive data collected from 1,500 initially recruited Wisconsin state employees. These employees were randomly selected to represent a wide demographic and health spectrum, aged 30–60 at the time of recruitment, ensuring a broad base for observing the natural history of sleep disorders and associated comorbidities such as CVDs. The WSCS encompasses 360 variables across multiple domains: anthropometry, clinical data, demographics, general health, lifestyle and behavioral health, sleep monitoring, sleep questionnaires, medical history, and sleep treatment. Each participant in the study was re-assessed at four-year intervals, with additional follow-ups to capture the progression and any new onset of health conditions, specifically focusing on the development and trajectory of CVD and its correlation with OSA.

Building on the expansive WSC database, our longitudinal analysis harnesses a refined dataset that reflects the progression of OSA-CVD comorbidity, in which 230 variables were shortlisted from the original 360 variables. The original design of the WSC planned for up to five follow-up visits per participant; however, the actual frequency and timing of these visits showed significant variability. Initial participation was robust with 1,123 participants at Visit 1, but subsequent visits saw a noticeable decline—748 participants at Visit 2, 566 at Visit 3, 121 at Visit 4, and only 2 at Visit 5. This attrition and irregularity in follow-up compliance highlight the challenges in longitudinal health research. Our study strategically narrows its focus to the first three visits, which not only boast higher participation rates but also offer more reliable data continuity, critical for assessing the longitudinal impact of sleep disorders on cardiovascular health. Furthermore, the intervals between visits varied widely, from 1 to 11 years, posing additional challenges for consistent data analysis. To address this, our analysis specifically targets 360 subjects who maintained regular follow-up intervals of 3, 4, or 5 years. By focusing on this subset, we aim to provide a detailed examination of the temporal relationships between OSA and CVD health outcomes, leveraging the strength of consistent and comprehensive data collection over time. **Table 1** provides a structured overview of the sequence of CVD outcomes over three successive visits, based on the *"any_cvd"* variable, which represents conditions such as heart attacks, congestive heart failure, and surgeries related to cardiovascular interventions.

**Table 1**. Longitudinal tracking of self-reported CVD outcomes in the WSCS database

| Group | Visit 1 (CVD Outcome) | Visit 2 (CVD Outcome) | Visit 3 (CVD Outcome) | Number of Patients |
|---|---|---|---|---|
| 0 | No CVD | No CVD | No CVD | 303 |
| 1 | No CVD | No CVD | CVD | 13 |
| 2 | No CVD | CVD | No CVD | 1 |
| 3 | No CVD | CVD | CVD | 9 |
| 4 | CVD | No CVD | No CVD | 4 |
| 5 | CVD | No CVD | CVD | 1 |
| 6 | CVD | CVD | No CVD | 0 |
| 7 | CVD | CVD | CVD | 29 |

### B. Data Preprocessing and Feature Engineering

Our data preprocessing involves several key procedures: (1) categorical encoding, (2) K-Nearest Neighbor (KNN)-based data imputation, (3) outliner detection, and (4) minority group over-sampling to address class imbalance. First, we transform categorical variables into a numerical format using a label-encoding algorithm. This conversion assigns a unique integer to each category, making the data suitable for the analytical models that require numerical input. To address missing data within our dataset, we implement a KNN-based imputation method [43]. This technique estimates and replaces missing values by analyzing the similarities between data instances. To identify and analyze the outliers in our high-dimensional dataset, we employ a multivariate outlier detection approach using the Mahalanobis distance [44]. This method calculates the distance of a point from the dataset's mean, scaled by the covariance among variables, making it ideal for identifying outliers in complex datasets. Once identified, outliers can be managed by removal or capping, depending on their impact on the analysis. Lastly, to enhance the LGMM model's performance and address the class imbalance issue within our dataset, we utilized the Synthetic Minority Over-sampling Technique (SMOTE) [45], this algorithm can generate new instances that are interpolations of the minority class samples that lie close together in the feature space. Here, we opt for a 100% upsampling of the CVD minority class, this is important for ensuring that less common but clinically significant patterns of CVDs are not overlooked, thereby improving the overall reliability and utility of our predictive LGMM model.

In refining our feature engineering methodologies for the multi-level phenotypic model of CVD and OSA, we employ a tree-based method including Random Forests or Gradient Boosting Machines that inherently evaluates the importance of each feature by measuring the reduction in impurity (typically Gini impurity or entropy) that each feature brings to the model when it is used in the tree construction [46]. Features that lead to significant splits in the tree are considered more important because they provide substantial information gain about the outcome variable. This process helps us pinpoint which clinical, demographic, and physiological features are most crucial in predicting the progression of CVD in patients with OSA. This approach ensures that the modeling focuses on the most relevant features, reducing the risk of overfitting and enhancing the generalizability of the model. Additionally, this method can uncover lesser-known or unexpected features that might play a significant role in CVD progression. The outputs from these algorithms will guide our subsequent modeling efforts, ensuring that our predictive models are not only robust but also tailored to highlight the nuances of CVD progression in the context of OSA.

### C. Multi-level Phenotypic Models of OSA-CVD Comorbidity

To investigate the complex interplay between OSA and CVDs, the LGMM model is developed to account for both fixed and random effects within longitudinal data. This model is specifically designed to handle variability both within and across patient individuals over time for studying the complex interactions and transitions among patients with varying CVD profiles. To further refine our analysis and enhance the understanding of patient trajectories, we employ t-SNE for dimensionality reduction paired with GMM for clustering. This innovative approach allows us to visualize and categorize complex phenotypes within our data, grouping patients into distinct clusters that represent varying risks and progression patterns of CVD. This step is crucial for identifying unique phenotypic expressions of CVD among OSA patients, enabling targeted interventions and personalized treatment strategies based on cluster-specific characteristics.

#### 1. Logistic mixed-effects model (LGMM)

In our study, the LGMM model is developed to dissect the complex interplay between OSA and CVDs, focusing on longitudinal patient data. This model is particularly adept at handling the intrinsic variations across individual patient trajectories and the recurring assessments inherent to our dataset, thus providing robust insights into the progression of CVD among OSA patients. The LGMM framework is designed to accommodate both fixed effects, which capture the general impact of observed variables across the entire population, and random effects, which allow for individual-specific variations. This dual approach is crucial in our study as it addresses both the common and unique factors contributing to CVD outcomes among OSA sufferers. For instance, the fixed components of our model assess the average influence of known risk factors like age, BMI, and smoking status on CVD incidence, while the random effects account for unobserved heterogeneity, potentially linked to genetic predispositions or lifestyle discrepancies. Mathematically, the LGMM for our OSA-CVD analysis is structured as follows:

$$\log\left(\frac{p_{ij}}{1-p_{ij}}\right) = \beta_0 + \mu_i + \beta_1 X_{1ij} + \beta_2 X_{2ij} + \cdots + \beta_n X_{nij} + \varepsilon_{ij} \qquad (1)$$

where:
- $p_{ij}$ is the probability of observing a cardiovascular event for the $i$-th subject at the $j$-th time point.
- $\beta_0$ represents the global intercept reflecting the baseline log odds of experiencing a CVD event.

- $\beta_k$ are the coefficients for each predictor $X_{kij}$, indicating the impact of both OSA-related factors (*e.g.,* severity of nocturnal hypoxia, sleep fragmentation) and general health indicators (*e.g.,* cholesterol levels, blood pressure, and lifestyle factors).
- $\mu_i$ denotes the random intercept for the $i$-th subject, capturing the unobserved heterogeneity due to individual susceptibility and resilience, which might influence the progression and outcomes of CVDs. These random effects are assumed to follow a normal distribution, $\mu_i \sim \mathcal{N}(0, \sigma_\mu^2)$.
- $\varepsilon_{ij}$ is the error term, accounting for the residual variability not explained by the model.

*2. High-dimensional patient trajectory tracking and phenotypic clustering with t-SNE and GMM*

Our study employs the t-SNE and GMM techniques not only to reduce the dimensionality of complex medical data but also to uncover latent structures and phenotypes that characterize the progression and variability of CVD among OSA patients. The t-SNE approach [47] provides a sophisticated means of visualizing the high-dimensional space of patient trajectories over multiple visits. Initially, t-SNE effectively reduces the dimensions of our dataset, which includes a broad range of clinical, biochemical, and demographic variables, into a two-dimensional space for visualization. Moreover, unlike other dimensionality reduction techniques, t-SNE preserves the local structure of the data, ensuring that similar data points in the high-dimensional space remain close in the reduced space. This characteristic is essential for accurately tracking the progression of CVD in patients with OSA. We configure t-SNE with a perplexity parameter tuned to the size of our dataset, optimizing the balance between the local and global aspects.

After reducing dimensionality, we employ the GMM to identify distinct phenotypic clusters within the t-SNE transformed data. GMM provides a probabilistic means to model the data points as a mixture of multiple Gaussian distributions. Each component of the mixture represents a potential phenotype of OSA-CVD comorbidity, allowing us to classify patients into distinct groups based on their clinical profiles. GMM accommodates the heterogeneity within our patient data, modeling the distribution of each cluster with its own covariance structure. This flexibility allows for the effective handling of the diverse manifestations of CVD in OSA patients, ranging from mild to severe forms. We utilize the EM algorithm to iteratively estimate the parameters of GMM, maximizing the likelihood function:

$$\boldsymbol{\theta} = \arg\max_{\boldsymbol{\theta}} \sum_{i=1}^{N} \log\left(\sum_{k=1}^{K} \pi_k \mathcal{N}(x_i | \mu_k, \boldsymbol{\Sigma}_k)\right) \quad (2)$$

where $\boldsymbol{\theta}$ represents the set of all parameters to be optimized in the GMM. These parameters include: (1) $\mu_k$: mean of the $k$-th Gaussian component, (2) $\boldsymbol{\Sigma}_k$: covariance matrix of the $k$-th Gaussian component, and (3) $\pi_k$ mixing coefficient (or weight) of the $k$-th Gaussian component, representing the probability that a randomly selected datapoint belongs to this component. To further refine and validate our phenotypic clusters, we apply the Kullback-Leibler Divergence (KLD) [48], also known as relative entropy, to measure the distinctiveness of each phenotypic cluster's distribution compared to another one. This step ensures that the clusters identified are statistically significant and not artifacts of the dimensionality reduction process. The KLD between two continuous distributions, $P$ and $Q$, with probabiltiy density functions $p(x)$ and $q(x)$:

$$D_{KL}(P\|Q) = \int_{-\infty}^{\infty} p(x) \log\left(\frac{p(x)}{q(x)}\right) dx \qquad (3)$$

### D. Model Validation

To rigorously assess the predictive performance of our models, we employ a comprehensive suite of statistical metrics including Accuracy ($Acc$), Precision ($Pre$), Specificity ($Spec$), Negative Predictive Value ($NPV$), and Recall. Accuracy measures the overall effectiveness of the model in correctly predicting both CVD and non-CVD cases, providing a general assessment of model performance. It is calculated as the ratio of correctly predicted observations (true positives ($TP$) and true negatives ($TN$)) to the total number of cases, given by the formula: $Acc = (TP + TN)/(TP + TN + FP + FN)$. Here, $FP$ and $FN$ are false positives and false negatives, respectively. Precision reflects the model's effectiveness in identifying only relevant instances among those retrieved. It is defined as the ratio of true positive predictions to all positive predictions: $Pre = TP/(TP + FP)$. Specificity ($Spec$) or the true negative rate, measures the proportion of actual negatives (non-CVD) correctly identified, an essential metric for confirming the absence of disease: $Spec = TN/(TN + FP)$. Negative Predictive Value ($NPV$) indicates the likelihood that subjects not predicted to have CVD truly do not have the disease, enhancing confidence in model predictions of non-occurrence: $NPV = TN/(TN + FN)$. Recall, also known as sensitivity or the true positive rate ($TPR$), quantifies the model's ability to identify all relevant instances (actual CVD cases), crucial for ensuring that all potential CVD cases receive appropriate clinical attention: $TPR = TP/(TP + FN)$. The effectiveness of the models will be further evaluated through 5-fold cross-validation and clinical expert knowledge, ensuring robustness and generalizability of the results across different subsets of data.

## III. RESULTS

This section presents the detailed outcomes of our analysis, systematically unfolding the complex interplay between OSA and CVDs within the WSCS. First, we performed descriptive analysis of variables across three visits and then evaluated the predictive power of various clinical indicators, using tree-based feature importance analysis to highlight key biomarkers that significantly influenced the CVD progression. Subsequently, the LGMM performance was evaluated, demonstrating its capability to capture longitudinal transitions and provide robust statistical validation. Next, we explored phenotypic clustering implemented by the t-SNE and GMM approaches, which revealed distinct CVD phenotypes through sophisticated dimensional reduction and clustering techniques. The validity of these phenotypic clusters was assessed using KLD, ensuring their distinctiveness and relevance in clinical contexts. Finally, comprehensive model validation metrics underscored the precision and reliability of our predictive LGMM.

### A. Comparative Analysis of Health Metrics Categorized by CVD Outcomes Across Three Visits

**Table 2** presents a comprehensive comparison of various health metrics categorized by CVD status across three successive visits within the WSCS. CVDs are classified into four MACEs based on the self-reported presence of specific conditions and treatments: MACE1 involves acute coronary syndrome/ischemic heart disease identified by factors such as heart attack and coronary artery conditions, MACE2 pertains to chronic heart failure, and MACE3 includes arrhythmias with treatments like pacemakers. The fourth MACE type, specifically detailed as treatment outcomes, further delineates the dataset. The table displays data for 360 participants, split into those with and without CVD at each visit, providing a vital understanding of the progression and potential remission of cardiovascular conditions. Notably, significant differences are observed in metrics such as cholesterol levels, hypertension, and diabetes medication usage, suggesting their strong association with CVD progression in patients with sleep apnea.

**Table 2**. Comparative analysis of variables by CVD outcomes across three visits.

|  | Visit 1 | | Visit 2 | | Visit 3 | |
| --- | --- | --- | --- | --- | --- | --- |
|  | No CVD (n = 326) | CVD (n = 34) | No CVD (n = 321) | CVD (n = 39) | No CVD (n = 308) | CVD (n = 52) |
| **Anthropometry, mean(SD)** | | | | | | |
| bmi | 31.35(6.79) | 31.64(6.64) | 31.43(7.22) | 31.91(6.35) | 31.15(7.12) | 32.22(7.07) |
| waisthip | 0.9(0.09) | 0.97(0.1) | 0.91(0.1) | 0.97(0.08) | 0.93(0.09) | 0.98(0.1) |
| hipgirthm | 109.92 (14.13) | 106.48 (11.57) | 109.83 (14.77) | 107.51 (12.56) | 110.37 (14.66) | 107.85 (13.74) |
| neckgirthm | 38.75(4.18) | 40.75(4.47) | 38.36(4.09) | 41.07(4.08) | 38.15(3.97) | 40.25(4.52) |
| headcm | 56.73(2.41) | 57.46(2.18) | 56.62(2.08) | 57.94(1.55) | 56.58(2.02) | 57.63(2.02) |
| **Clinical Data, mean(SD)** | | | | | | |
| total_cholesterol | 203.16 (31.96) | 176.97 (32.35) | 197.97 (36.02) | 156.79 (22.09) | 187.25 (33.58) | 152.69 (35.7) |
| ldl | 121.64 (29.05) | 95.65 (31.73) | 114.96 (32.3) | 81.1 (22.83) | 107.49 (29.36) | 78.04 (26.29) |
| creatinine | 1 (0.18) | 1.03 (0.18) | 1 (0.2) | 1.06 (0.2) | 0.93 (0.24) | 1.05 (0.25) |
| triglycerides | 146.21 (80.75) | 160.24 (74.93) | 133.73 (78.59) | 127.15 (72.85) | 138.51 (74.91) | 149.56 (85.08) |
| uric_acid | 5.69 (1.35) | 6.05 (1.53) | 5.74(1.32) | 6.37 (1.33) | 5.1 (1.17) | 5.84 (1.43) |
| sbp_mean | 125.47 (14.6) | 131.85 (12.34) | 126.34 (15.28) | 127.05 (13.85) | 127.07 (1.17) | 128.08 (1.43) |
| hdl | 52.88 (14.8) | 49.18 (15.00) | 57.05 (15.84) | 51.81 (14.58) | 52.83 (16.62) | 45.79 (13.58) |
| glucose | 103.31 | 120.82 | 106.17 | 109.08 | 101.26 | 100.01 |
| | (23.85) | (49.33) | (28.59) | (20.03) | (18.8) | (19.73) |
| **Demographics** | | | | | | |
| sex, n(%) | | | | | | |
| *F* | 170(52.15) | 26(76.47) | 164(51.09) | 32(82.05) | 157(50.97) | 39(75) |
| *M* | 156(47.85) | 8(23.53) | 157(48.91) | 7(17.95) | 151(49.03) | 13(33.33) |
| age, mean (SD) | 55.97 | 60.82 | 59.94 | 65.05 | 64.36 | 68 |
| | (7.38) | (6.91) | (7.08) | (7.23) | (6.98) | (7.75) |
| **General Health, n(%)** | | | | | | |
| eval_heath | | | | | | |
| *1: Excellent* | 52(15.95) | 5(14.71) | 51(15.89) | 3(7.69) | 40(12.99) | 6(11.54) |
| *2: Very good* | 163(50) | 10(29.41) | 157(48.91) | 11(28.21) | 14(48.38) | 10(19.23) |
| *3: Good* | 92(28.22) | 19(55.88) | 92(28.66) | 21(53.85) | 95(30.84) | 30(57.69) |
| *4: Fair* | 18(5.52) | N/A | 20(6.23) | 4(10.26) | 20(6.49) | 5(9.62) |
| *5: Poor* | 1(0.31) | N/A | 1(0.31) | N/A | 4(1.3) | 1(1.92) |
| **Lifestyle and Behavioral Health, n(%)** | | | | | | |
| caffeine | 2.74(2.23) | 3.06(2.77) | 2.57(2.16) | 2.67(2.17) | 2.56(2.05) | 3.1(2.45) |
| MACEs | | | | | | |
| *MACE1* | N/A | 29(85.29) | N/A | 34(87.18) | N/A | 45(86.54) |
| *treatMACE1* | N/A | 23(67.65) | N/A | 33(84.62) | N/A | 36(69.23) |
| *MACE3* | 41(12.58) | 14(41.18) | 37(11.53) | 17(43.59) | 39(12.66) | 17(32.69) |
| *treatMACE3* | N/A | 1(2.94) | N/A | 3(7.69) | N/A | 6(11.54) |
| *MACE2* | N/A | 1(2.94) | N/A | 4(10.26) | N/A | 3(5.77) |
| *MACE4* | 4(1.23) | 1(2.94) | 7(2.18) | 2(5.13) | 6(1.95%) | 3(5.77) |
| **Medical History, n(%)** | | | | | | |
| narcotics_med | 6(1.84) | 1(2.94) | 4(1.25) | 3(7.69) | 6(1.95) | 2(3.85) |
| anxiety_med | 26(7.98) | N/A | 22(6.85) | 1(2.56) | 23(7.47) | 4(7.69) |
| sedative_med | 25(7.67) | 2(5.88) | 29(9.03) | 4(10.26) | 39(12.66) | 11(21.15) |
| htn_med | 105(32.21) | 25(73.53) | 131(40.81) | 35(89.74) | 147(47.73) | 48(92.31) |
| arthritis_ynd | 95(29.14) | 13(38.24) | 113(35.2) | 16(41.03) | 129(41.88) | 27(51.92) |
| diabetes_med | 17(5.21) | 6(17.65) | 28(8.72) | 11(28.21) | 41(13.31) | 17(32.69) |
| **Sleep Monitoring, mean(SD)** | | | | | | |
| nremahi | 8.49 | 13.16 | 10.19 | 14.39 | 8.35 | 14.51 |
| | (14.04) | (16.97) | (13.69) | (17.51) | (12.48) | (17.74) |
| ahi | 10.39 | 15.21 | 12.23 | 16.61 | 10.03 | 16.4 |
| | (14.56) | (18.05) | (14.29) | (17.9) | (13.1) | (18.35) |
| avgo2sattst | 95.33(1.53) | 95.2(1.82) | 95.41(3.02) | 96.45(7.13) | 95.46(3.69) | 95.27(6.49) |
| pcttstrem | 17.11(5.94) | 16.66(5.71) | 16.47(5.88) | 15.85(6.65) | 15.15(5.96) | 14.3(6.72) |
| sleep_latency | 13.77 | 9.68 | 12.05 | 10.95 | 16.07 | 12.51 |
| | (17.32) | (11.9) | (13.08) | (17.57) | (18.12) | (13.49) |
| **Sleep Questionnaires, n(%)** | | | | | | |
| apnea | 32(9.82) | 4(11.76) | 46(14.33) | 11(28.21) | 65(21.1) | 18(34.62) |
| **Sleep Treatment** | | | | | | |
| apnea_treatment | 30(9.2) | 4(11.76) | 46(14.33) | 11(28.21) | 64(20.78) | 16(30.77) |

As shown in **Table 2**, it revealed a pattern of gradual health deterioration among CVD patients in comparison with non-CVD subjects, marked by specific metrics in anthropometry, clinical data, and general health assessments. Across the three visits, body mass index (BMI) in the CVD group showed a slight increasing trend from 31.64 to 32.22, compared to a marginal decrease in the non-CVD group from 31.35 to 31.15. This suggests a potential worsening of body composition in CVD patients. Similar trends were observed in the waist-hip ratio, increasing from 0.97 to 0.98 in CVD patients, while it remained relatively stable in the non-CVD group. These metrics indicate an increasing risk factor for cardiovascular complications due to changes in body composition. As for clinical data, there is a notable decline in total cholesterol levels across the three visits for individuals with CVD, more pronounced than in those without CVD. From Visit 1 to Visit 3, total cholesterol dropped from 176.97 mg/dL to 152.69 mg/dL in the CVD group, compared to a decrease from 203.16 mg/dL to 187.25 mg/dL in the non-CVD group. This might reflect the effect of targeted cholesterol-lowering therapies or the natural progression of the disease. Similar to total cholesterol, low-density lipoprotein (*"ldl"*) cholesterol levels decreased more significantly in the CVD group, from 95.65 mg/dL to 78.04 mg/dL, indicating effective management or the impact of disease progression. However, despite these improvements, the CVD group exhibited higher systolic blood pressure (*"sbp_mean"*) across the visits, slightly increasing to 128.08, which could indicate persistent challenges in blood pressure management.

Regarding general health and demographics, the analysis showed a higher proportion of females in the CVD group, which is significant given the potential gender differences in cardiovascular risk and outcomes. The age profile also indicated that CVD patients were generally older, with ages increasing from 60.82 to 68 by the third visit, underscoring age as a significant cardiovascular risk factor. In terms of medical interventions and health assessments, there was a notable escalation in the incidence of MACEs across the visits, reflecting the progression of the disease and the need for increased medical interventions. The data shows a clear increase in the use of treatments for conditions identified as MACE1 and MACE3, indicating a response to worsening cardiovascular symptoms. Sleep-related measures such as the apnea-hypopnea index (AHI) worsened over time in CVD patients, suggesting a possible link between deteriorating sleep quality and CVD health. Sleep apnea treatments were more prevalent among the CVD group, aligning with the observed increase in sleep disorder severity. The use of diabetes medications is also higher in the CVD group (17.65% in Visit 1 to 32.69% by Visit 3), reflecting the common comorbidity of diabetes with CVDs

### B. Feature Importance Analysis

Our analysis utilized a tree-based method, specifically Random Forests [49], to rank the top 20 features across the WSCS dataset, integrating SMOTE and 5-fold cross-validation to enhance the robustness and reliability of the results. Here, we employed SMOTE to address the significant class imbalance present in our dataset, where instances of non-CVD outcomes heavily outnumbered CVD outcomes. The SMOTE algorithm was configured to oversample the minority class (CVD cases) by 100%, effectively doubling the number of CVD cases in the training data. This oversampling was critical to provide a more balanced dataset, which helps in improving the classifier's ability to detect the minority class without overfitting. The Random Forest algorithm was utilized for feature ranking due to its efficacy in handling high-dimensional data and its robustness against overfitting. We configured the Random Forest with 100 trees, ensuring a comprehensive exploration of the feature space. Each tree in the forest was allowed to grow to its maximum length without pruning, which permits the model to learn highly detailed patterns in the data. Feature importance was then derived from the average decrease in Gini impurity across all trees when splitting on a particular feature, providing a robust measure of each feature's predictive power. The ranked features, as presented in **Figure 2**, revealed a pronounced emphasis on clinical data, with total cholesterol, LDL cholesterol, and diabetes medications scoring the highest in importance.

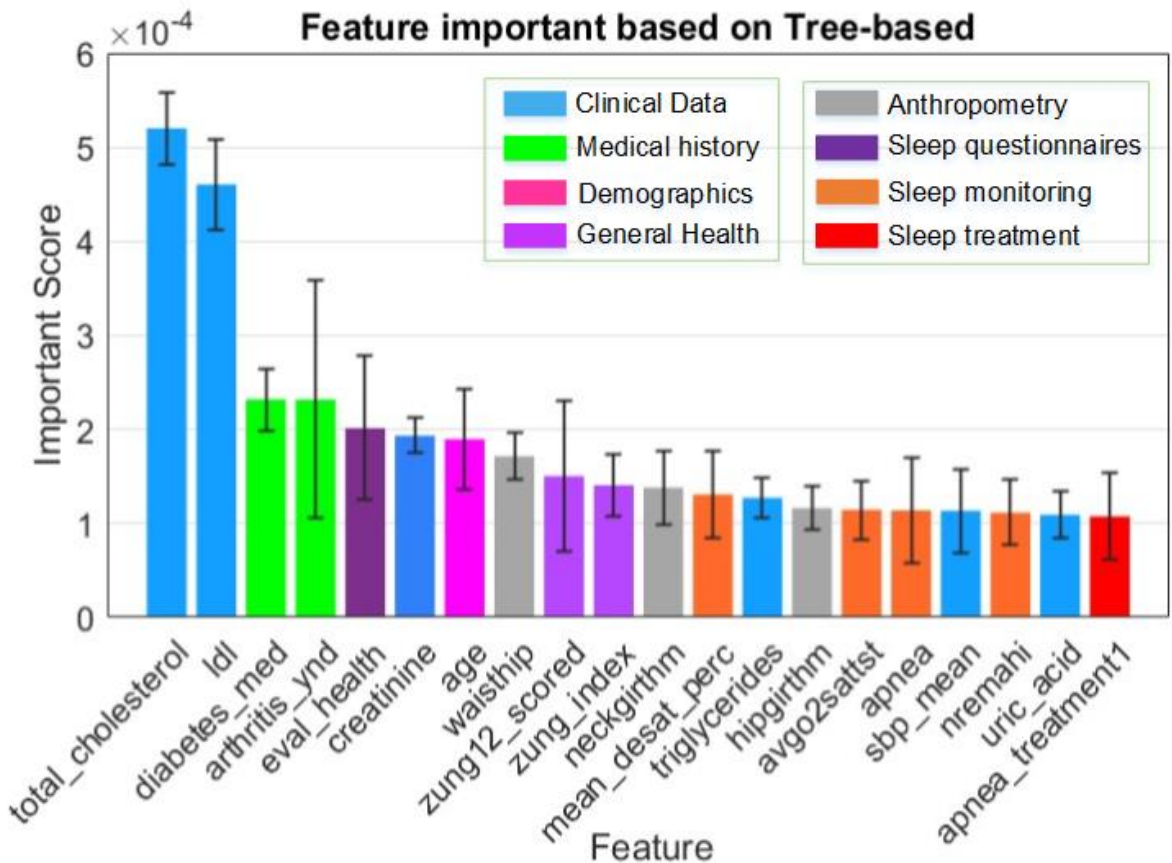

**Figure 2**. Feature importance scores from tree-based model analysis. This graph quantifies the predictive significance of each feature, with colors representing different data categories (*e.g.,* clinical data, medical history, demographics, and general health).

According to **Figure 2**, medical history variables such as arthritis medication use and health evaluations also showed significant relevance, reflecting their contribution to the nuanced prediction of CVD risks. Notably, general health measures, including serum creatinine levels, and demographic features such as age and waist circumference, were identified as key predictors. Furthermore, metrics from sleep monitoring—specifically the average level of oxygen desaturation and mean oxygen saturation during total sleep—emphasized the interconnections between sleep quality and cardiovascular health. The integration of anthropometric measures and Zung's self-rating depression scale scores illustrated the model's capability to capture the broad spectrum of factors influencing CVD risk.

## C. Logistic Mixed-Effects Model Validation

Our evaluation of the LGMM extensively assessed its performance in predicting CVD outcomes **(see Figure 3)**. This assessment utilized the top 20 features identified through a tree-based method, revealing significant model performance distinctions between LGMM and logistic regression (LR).

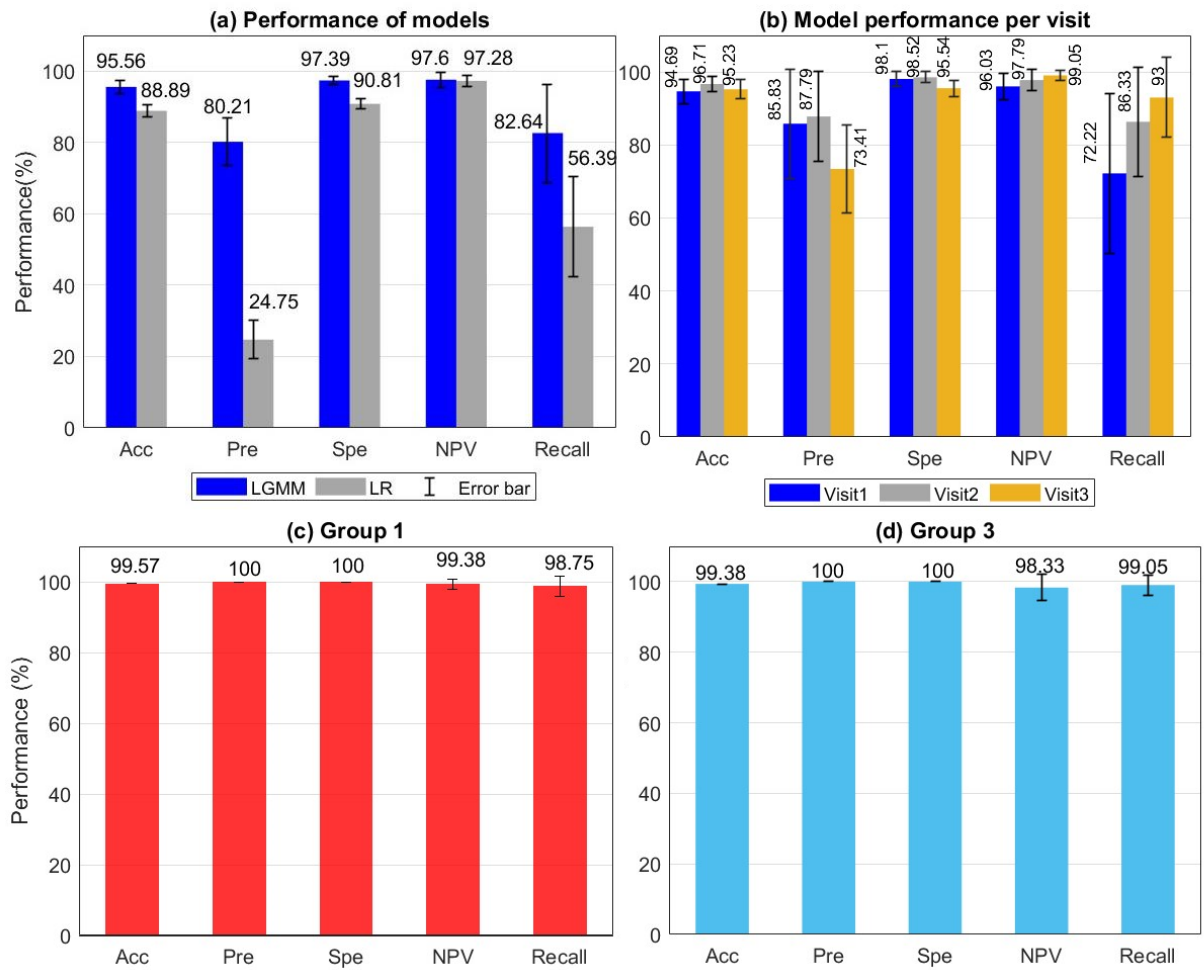

**Figure 3**. Comprehensive evaluation of LGMM's performance across different scenarios. Panel (a) showcases the comparative performance of LGMM and LR models. Panel (b) details LGMM's performance across three visits. Panels (c) and (d) depict LGMM's robustness in patient group 1 and group 3.

Here, an independent LR model was developed for each visit (Visit 1, Visit 2, Visit 3). Each model utilized the same top 20 features identified through the tree-based method that was employed for the LGMM. This ensured that any observed differences in model performance were due to the model's structure and handling of the data, rather than differences in the input variables. The LGMM and LR were trained on a dataset processed through 5-fold cross-validation, a method chosen to mitigate overfitting and ensure the generalizability of the model predictions. **Figure 3(a)** details the aggregate performance metrics of both models across all data points pooled from the three visits. The LGMM demonstrated superior accuracy at 95.88%, which significantly surpassed the LR's 82.64%. In terms of precision, the LGMM achieved 88.90% compared to LR's significantly lower 24.75%. This disparity highlights the LGMM's enhanced ability to correctly identify positive CVD cases. The recall metric further emphasizes the LGMM's robustness, with a score of 97.28%, indicating that it successfully identified nearly all actual CVD cases, whereas the LR managed only 56.39%.

**Figure 3(b)** presents a detailed evaluation of the LGMM across three sequential visits within the study, revealing the model's sustained effectiveness in predicting CVD outcomes over time. The accuracy of the model consistently hovered around 95% for each visit. Specifically, the precision exhibited a modest decrease from 85.83% in the initial visit to approximately 73.41% in the visit 3, potentially indicating slight variations in the model's positive predictive capacity as patient conditions evolve. The specificity and NPV values remained exceptionally high, both averaging close to 98%, which indicated the model's consistent ability to correctly identify non-CVD cases across different visits. Recall, or the model's sensitivity, rose gradually from around 72.22% during the first visit to 93% by the third visit. This incremental improvement suggests that the LGMM becomes increasingly effective at capturing true CVD cases.

**Figure 3(c-d)** depicts the evaluation of the LGMM's performance for Group 1 and Group 3 from **Table 1**. The LGMM for each group was trained independently based on the corresponding group dataset, and the data was oversampled using SMOTE for both CVD and non-CVD classes by 500% for obtaining sufficient training data. This was followed by the tree-based method implementation to shortlist the top 10 features. For Group 1, the model achieves near-perfect scores, with 100% specificity and NPV, and 99.57% accuracy, suggesting an exceptional ability to identify and validate the onset of CVD conditions later in the timeline. In contrast, for Group 3, which deals with patients already showing symptoms by the second visit, the model also exhibits high accuracy (99.38%) and specificity (100%), with a slightly lower recall of 99.05%. This indicates that the model is equally capable of handling cases where the disease is more pronounced and ongoing, maintaining high precision in monitoring CVD progression. This detailed

subgroup analysis not only confirms the model's reliability and effectiveness in different patient groups but also underscores its potential in tailored patient monitoring, thereby enhancing the predictive framework for CVD outcomes based on individual patient trajectories and disease timelines.

In the detailed analysis of fixed effects within our LGMM, **Table 3** presents the estimated fixed effects of key variables alongside their statistical significance.

**Table 3**. Estimated fixed effects of the representative variables from the LGMM analysis.

| Variables | Estimate | SE | CI upper | CI lower | p-value | Sig code |
|---|---|---|---|---|---|---|
| wsc_vst | -0.0275 | 0.012 | 4.0797 | -1.9246 | 0.0224 | * |
| total_cholesterol | -0.0081 | 0.0027 | -0.0039 | -0.0511 | 0.0031 | ** |
| triglycerides | 0.0004 | 0.0004 | 0.0344 | -0.0184 | 0.245 | |
| sbp_mean | 0.0203 | 0.0062 | 0.3426 | -0.3433 | 0.001 | ** |
| hipgirthm | -0.013 | 0.0092 | 0.0037 | -0.0046 | 0.1583 | |
| zung_index | 0.0137 | 0.0077 | -0.0028 | -0.0135 | 0.0768 | + |
| diabetes_med | 0.0584 | 0.0345 | 0.0607 | -0.1032 | 0.0904 | + |
| nremahi | 0.0027 | 0.0016 | 0.0324 | 0.0082 | 0.0878 | + |
| avgo2sattst | -0.0246 | 0.0205 | 0.0051 | -0.0311 | 0.2306 | |
| mean_desat_perc | -0.0333 | 0.0218 | 1.5794 | -4.3209 | 0.1264 | |

*Significance codes: $p < 0.001$ '***', $p < 0.01$ '**', $p < 0.05$ '*', and $p < 0.1$ '+'*

As shown in **Table 3**, the variable *"wsc_vst"* (visit sequence) had an estimate of -0.0275 with a p-value of 0.0224, indicating a significant negative association with CVD outcomes, suggesting that later visits had a slightly lower risk of CVD, potentially reflecting some stabilization or effective management over time. Total cholesterol (*"total_cholesterol"*) showed a significant negative association with CVD outcomes (estimate: -0.0081, p-value: 0.0031). This negative relationship indicated that higher cholesterol levels correlated with a decreased probability of CVD occurrence, which might seem counterintuitive but may reflect the complexity of cholesterol metabolism in the OSA-CVD comorbidity. Systolic blood pressure (*"sbp_mean"*) showed a significant positive effect on CVD outcomes (estimate: 0.0203, p-value: 0.001), indicating that higher blood pressure was associated with increased CVD risk. Hip girth (*"hipgirthm"*) had an estimate of -0.013 with a p-value of 0.1583, indicating it was not statistically significant. Similarly, neck girth (*"nremahi"*) with an estimate of 0.0027 and a p-value of 0.0878, although approaching significance, suggested a potential positive association with CVD risk. The Zung self-rating depression scale index (*"zung_index"*) showed a borderline significant positive effect (estimate: 0.0137, p-value: 0.0768). Diabetes medication use (*"diabetes_med"*) was another important variable, showing a positive effect (estimate: 0.0584, p-value: 0.0904), indicating that individuals on diabetes medication had a higher risk of CVD, which aligned with the known comorbidity between diabetes and CVDs. For sleep-related variables, the average non-REM AHI (*"nremahi"*) showed a positive association with CVD risk (p-value: 0.0878), albeit marginally significant, underscoring the impact of OSA severity on cardiovascular health. However, other sleep metrics such as average oxygen saturation (*"avgo2sattst"*) and mean desaturation percentage (*"mean_desat_perc"*) did not show significant effects. In our detailed subgroup analyses, **Table 4** presents the estimated fixed effects for representative variables within Group 1 and Group 3.

**Table 4**. Estimated fixed effects of the representative variables for Group 1 and Group 3

| Variables | Estimate | SE | CI upper | CI lower | p-value | Sig code |
|---|---|---|---|---|---|---|
| **Group 1** | | | | | | |
| wsc_vst | 0.3678 | 0.0215 | 0.4102 | 0.3253 | < 0.0001 | *** |
| creatinine | 1.6643 | 0.6284 | 2.9046 | 0.424 | 0.0088 | ** |
| hdl | 0.0325 | 0.01 | 0.0522 | 0.0128 | 0.0013 | ** |
| ldl | -0.0113 | 0.0026 | -0.0062 | -0.0164 | < 0.0001 | *** |
| apnea_treatment1 | 0.06 | 0.0164 | 0.0923 | 0.0277 | 0.0003 | *** |
| anxiety_med | -0.2719 | 0.17 | 0.0636 | -0.6073 | 0.1115 | |
| sedative_med | 0.4033 | 0.1475 | 0.6944 | 0.1122 | 0.0069 | ** |
| sleep_latency | 0.0237 | 0.0065 | 0.0366 | 0.0108 | 0.0004 | *** |
| creatinine^2 | -0.9554 | 0.3134 | -0.3368 | -1.5741 | 0.0027 | ** |
| sleep_latency^2 | -0.0005 | 0.0002 | -0.0002 | -0.0009 | 0.005 | ** |
| **Group 3** | | | | | | |
| wsc_vst | 0.3758 | 0.0217 | 56.363 | -18.557 | < 0.0001 | *** |
| glucose | 0.0311 | 0.0063 | 0.4188 | 0.3327 | < 0.0001 | *** |
| ldl | -0.0092 | 0.0025 | 0.0435 | 0.0186 | 0.0004 | *** |
| headcm | -0.756 | 0.6527 | 0.0133 | -0.0202 | 0.2491 | |
| caffeine | 0.0693 | 0.0207 | -0.0042 | -0.0142 | 0.0011 | ** |
| pcttststage34 | 0.0841 | 0.013 | 0.5369 | -2.049 | < 0.0001 | *** |
| mean_desat_perc | 0.0544 | 0.0335 | 0.1103 | 0.0283 | 0.107 | |
| headcm^2 | 0.0069 | 0.0056 | -0.0001 | -0.0002 | 0.2261 | |
| caffeine^2 | -0.0111 | 0.0028 | 0.0002 | -0.0001 | 0.0001 | *** |
| pcttststage34^2 | -0.0043 | 0.001 | 0 | 0 | 0.0001 | *** |

*Significance codes: $p < 0.001$ '***', $p < 0.01$ '**', $p < 0.05$ '*', and $p < 0.1$ '+'*

As observed in **Table 4**, Group 1 displayed significant findings in several variables. The variable *"wsc_vst"* had an estimate of 0.3678 with a p-value of less than 0.001, indicating a strong positive association with CVD outcomes. This suggests that as the visits progressed, the likelihood of CVD diagnosis increased, reflecting the worsening health condition over time for these patients. Creatinine levels (*"creatinine"*) showed a significant positive association with CVD outcomes (estimate: 1.6643, p-value: 0.0088), indicating that higher creatinine levels, which often denote poorer kidney function, were linked to increased CVD risk. High-density lipoprotein (*"hdl"*) cholesterol also showed a positive association (estimate: 0.0325, p-value: 0.0013). Low-density lipoprotein (*"ldl"*) cholesterol was significantly negatively associated with CVD outcomes (estimate: -0.0113, p-value: <0.001), The use of apnea treatment (*"apnea_treatment1"*) was positively associated with CVD (estimate: 0.06, p-value: <0.001), indicating that those undergoing treatment for sleep apnea had a higher likelihood of CVD, perhaps due to the severity of their OSA condition necessitating treatment. Sedative medication (*"sedative_med"*) was positively associated with CVD (estimate: 0.4033, p-value: 0.0069), suggesting that sedative use might contribute to increased CVD risk, potentially due to its effects on sleep architecture. Sleep latency (*"sleep_latency"*) had a significant positive effect on CVD risk (estimate: 0.0237, p-value: 0.0004), indicating that longer time taken to fall asleep was associated with higher CVD risk. The quadratic term for creatinine (*"creatinine^2"*) showed a significant negative effect (estimate: -0.9554, p-value: 0.0027), highlighting a non-linear relationship where very high levels of creatinine might have different implications compared to moderate elevations. Similarly, the quadratic term for sleep latency (*"sleep_latency^2"*) also showed significance (estimate: -0.0005, p-value: 0.005), indicating a complex non-linear impact on CVD risk.

For Group 3, the *"wsc_vst"* variable had an estimate of 0.3758 with a p-value < 0.001. Glucose levels (*"glucose"*) were significantly positively associated with CVD outcomes (estimate: 0.0311, p-value: <0.001), indicating that higher glucose levels were linked to increased CVD risk. LDL cholesterol showed a significant negative association (estimate: -0.0092, p-value: 0.0004), similar to Group 1. Caffeine intake (*"caffeine"*) was positively associated with CVD risk (estimate: 0.0693, p-value: 0.0011). The percentage of total sleep time in stage 3 and 4 sleep (*"pcttststage34"*) was significantly associated with CVD outcomes (estimate: 0.0841, p-value: <0.001), suggesting that deeper stages of sleep had a notable impact on CVD risk. Mean desaturation percentage (*"mean_desat_perc"*) showed a positive but non-significant association (estimate: 0.0544, p-value: 0.107). Quadratic terms for head circumference (*"headcm^2"*), caffeine intake (*"caffeine^2"*), and sleep stages (*"pcttststage34^2"*) were included to capture non-linear effects. Caffeine intake squared ("caffeine^2") showed a significant negative association (estimate: -0.0111, p-value: 0.0001). The quadratic term for sleep stages ("pcttststage34^2") was also significant (estimate: -0.0043, p-value: 0.0001), further highlighting the nonlinear complexity of sleep architecture in CVD risk.

### D. Patient Trajectory Analysis of OSA-CVD Comorbidity and Phenotypic Clustering

In our investigation of OSA-CVD comorbidity and phenotypic clustering, we concentrated on groups 1, 3, and 7 from **Table 1** due to their distinct CVD development stages and their predominant sample sizes. We started by applying t-SNE to reduce the dimensionality of the dataset, which includes the top 20 features (illustrated in **Figure 2**) identified as the most predictive of CVD outcomes. The t-SNE algorithm was configured with a perplexity setting of 30, aiming to balance local and global aspects of the data, and an early exaggeration factor of 12 to ensure distinct clustering of the data points. We utilized the Euclidean distance to measure the similarity between data points. Following the dimensionality reduction achieved with t-SNE, we employed GMM for clustering the reduced dataset. We determined the optimal number of clusters by evaluating the Bayesian Information Criterion (BIC) [50], ultimately choosing two clusters that corresponded to the natural grouping suggested by t-SNE outputs. We set "full" covariance type in GMM to allow each cluster its own general covariance matrix. The EM algorithm was used to estimate the model parameters. Utilizing the t-SNE and GMM approaches, we explored the patient transitions of these three groups across three clinical visits, as shown in **Figure 4**. This approach enabled us to track the progression and cluster formation based on phenotypic characteristics observed over time.

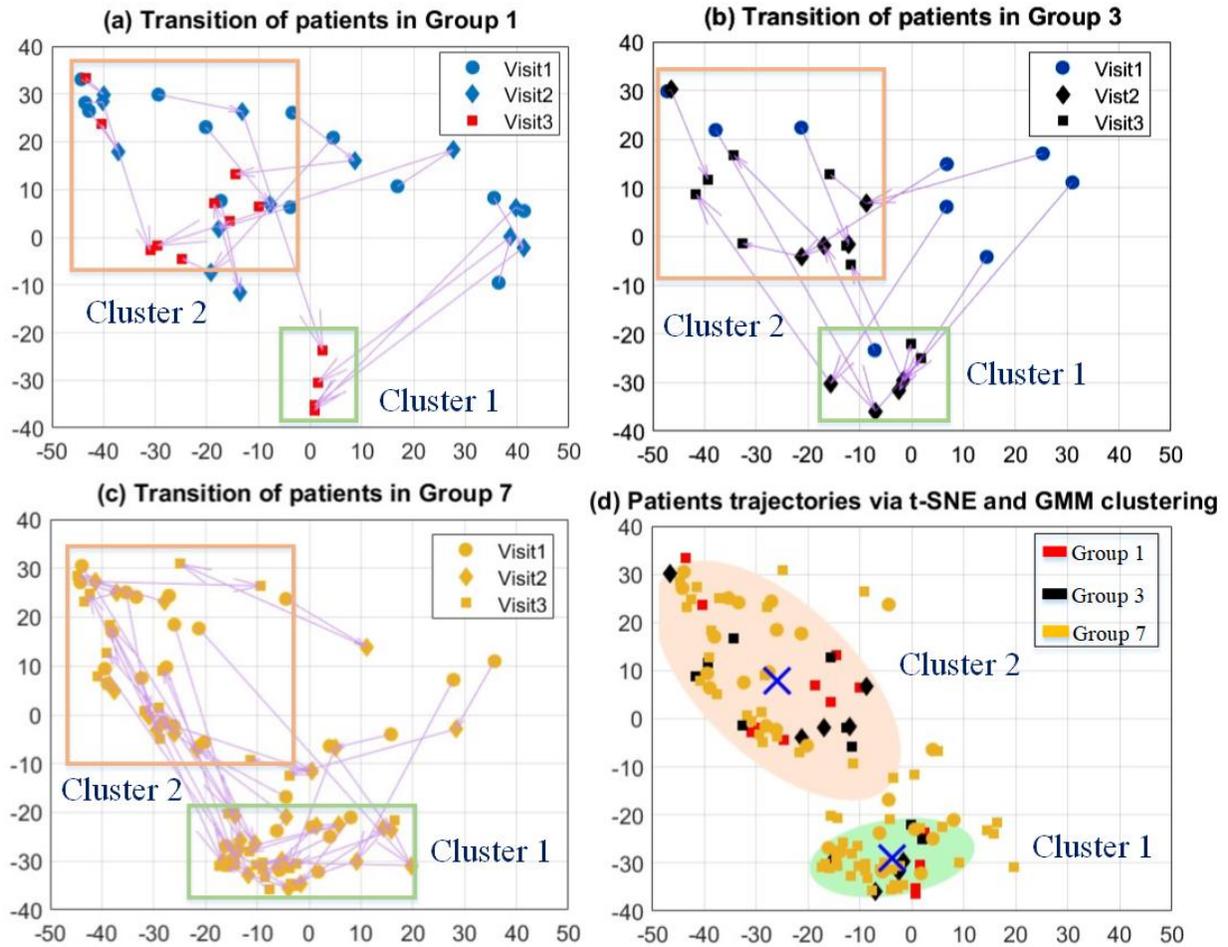

**Figure 4**. Patient transition and trajectories across three groups using t-SNE and GMM clustering. Panels (a)-(c) track the transitions of patients in Groups 1, 3, and 7 across three visits, highlighting the clustering dynamics. Panel (d) visually represents the clustering outcomes by group and visit.

In **Figure 4(a)** for group 1, the transition of patients across three visits spanning 6-10 years shows them initially dispersed but progressively converging into two distinct clusters. By the third visit, these clusters are well-defined, with patients in Cluster 1 displaying a consistent downward transition towards CVD manifestation compared to more complex trajectories in Cluster 2. This consistent transition in Cluster 1 might suggest either a more effective management of risk factors or a later onset of critical CVD symptoms. Furthermore, the average Euclidean distances between visits showed remarkable variations, especially between the second and third visits. Specifically, in Cluster 1, the distance between the first and second visits was relatively short at 10.047 units, indicating minor changes during this period. However, the distance dramatically increased to 52.61 units from the second to the third visit, reflecting a pronounced transition to a CVD-onset state. Conversely, Cluster 2 displayed more consistent distances: 15.47 units from the first to the second visit and a slightly increased distance of 17.6 units from the second to the third visit, suggesting a more gradual progression of CVD.

For Group 3 (**Figure 4(b)**), the trajectory analysis from visit 1 to visit 3 shows a rapid convergence into two clusters by the second visit, reflecting a faster progression of CVD. This group's patients move from a pre-CVD state at visit 1 to a definite CVD state by visit 2. Those transitioning to Cluster 2 by visit 2 likely exhibit more severe comorbidities or a higher accumulation of risk factors, leading to deteriorated health outcomes by visit 3. This transition was quantified by a substantial average Euclidian distance of 34.71 units between the first and second visits, after which the transition to the third visit was less pronounced, at 18.19 units. The trajectory data suggest that patients in Cluster 1 of this group maintain a relatively healthier profile, while those in Cluster 2 develop more severe comorbidities, as evidenced by shorter distances indicating a quicker convergence to adverse health outcomes.

For Group 7 observed in **Figure 4(c)**, the t-SNE visualization reveals more complex dynamics than initially described, showing notable transitions both between and within clusters over the three visits. This group, characterized by long-term CVDs, exhibited significant within-cluster movement as well as shifts from one cluster to another. From Visit 1 to Visit 2, several patients in Cluster 1 move towards and even into Cluster 2, suggesting a change in their health status or a progression in disease severity. This movement continues into Visit 3, where additional transitions are observed not just into but also within Cluster 2, indicating ongoing changes in patient conditions. The average Euclidean distances corresponding to intra-cluster transitions were substantially lower compared to the other groups as opposed to substantial inter-cluster transition distances due to dramatical changes in CVD progression. This intra- and inter-cluster mobility may reflect fluctuations in individual patient health parameters, potentially driven by variations in treatment efficacy, changes in lifestyle, or the natural progression of their disease. The distance metrics between visits within clusters further suggest variability in how CVD impacts these patients over time.

In panel (d) of **Figure 4**, the patients' trajectories across Groups 1, 3, and 7 were visualized using t-SNE combined with GMM clustering, which distinctly categorized these trajectories into two primary clusters. Cluster 1, depicted in a lighter shade, primarily comprised patients from Group 7 across all visits, indicating a consistent clustering of long-term CVD patients who exhibited stable yet distinct health characteristics compared to other groups. This cluster also included patients from Groups 1 and 3, primarily in their earlier visits, suggesting these patients' initial health states were similar to those of long-term CVD patients. Cluster 2, shown in a darker shade, mainly consisted of patients from Group 7 and Group 3 who developed CVD earlier, by the second visit, and Group 1 patients by their third visit. The representation in Cluster 2 implied a more aggressive or rapidly progressing form of CVD, as these patients transitioned to more severe states quicker than those in Cluster 1. The positioning of the markers (circle for Visit 1, diamond for Visit 2, square for Visit 3) within these clusters further delineated the trajectory of each patient group. Group 1 (Visit 3 prominently in Cluster 2) showed a clear progression from a less severe state in earlier visits to a more severe state by the third visit, indicating the late onset of CVD symptoms and their escalation. Group 3 transitioned earlier into Cluster 2 and stayed there, illustrating an earlier onset and consistent progression of CVD across subsequent visits. Group 7 remained in Cluster 1, underscoring their ongoing management of CVD without significant progression into more severe states as seen in Cluster 2. In the analysis detailed in **Table 5**, we evaluated the differences in the clinical, demographic, and sleep-related variables between the two clusters utilizing hypothesis testing to establish the statistical significance of observed differences.

**Table 5**. Statistical significance of the differences in MACE outcomes, anthropometric measures, clinical data, demographic factors, medical history, and sleep monitoring metrics between 2 clusters

| Categories | Variables | Cluster 1 (n = 49) | Cluster 2 (n = 54) | p-value | Sig code |
|---|---|---|---|---|---|
| MACE, n(%) | MACE1 | 47(95.92%) | 54(84.38%) | 0.2239 | |
| | MACE1 treatment | 45(91.84%) | 42(65.63%) | 0.0597 | + |
| | MACE3 | 18(36.73%) | 25(39.06%) | 0.4239 | |
| | MACE3 treatment | 3(6.12%) | 5(7.81%) | 0.7182 | |
| Anthropometry, mean(SD) | hipgirthm | 102(7.78) | 112.69(13.6) | < 0.001 | *** |
| | BMI | 29.08(4.41) | 34.77(6.77) | 0.6804 | |
| | waisthip | 0.99(0.09) | 0.98(0.09) | 0.0012 | ** |
| | neckgirthm | 40.41(3.41) | 41.24(4.48) | 0.0016 | ** |
| Clinical Data, mean(SD) | total_cholesterol | 141.12(17.93) | 167.81(31.12) | 0.1416 | |
| | ldl | 70.69(15.15) | 86.83(25.58) | 0.2632 | |
| | creatinine | 1.11(0.22) | 1.01(0.21) | < 0.0001 | *** |
| | triglycerides | 88.98(23.21) | 195.75(80.06) | < 0.0001 | *** |
| | sbp_mean | 131.97(13.22) | 127.05(12.5) | 0.1551 | |
| Demographics | sex: M, n(%) | 48(97.96%) | 22(34.38%) | < 0.0001 | *** |
| | age, mean(SD) | 67.73(6.64) | 64.39(8.07) | < 0.0001 | *** |
| Medical History, n(%) | narcotics_med | 2(4.08%) | 4(6.25%) | 0.0903 | + |
| | htn_diuretic_med | 12(24.49%) | 31(48.44%) | < 0.0001 | *** |
| | arthritis_ynd | 17(34.69%) | 36(56.25%) | 0.0621 | + |
| | diabetes_med | 12(24.49%) | 21(32.81%) | 0.1708 | |
| | apnea | 11(22.45%) | 18(28.13%) | 0.0423 | * |
| Sleep Monitoring, mean(SD) | nremahi | 11.76(12.46) | 16.98(20.84) | 0.2211 | |
| | ahi | 13.21(12.55) | 19.57(21.55) | 0.6180 | |
| | avgo2sattst | 94.99(3.52) | 96.13(7.51) | 0.1404 | |
| | pcttstrem | 15.02(6.87) | 15.2(6.37) | 0.4497 | |

*Significance codes: $p < 0.001$ '\*\*\*', $p < 0.01$ '\*\*', $p < 0.05$ '\*', and $p < 0.1$ '+'*

In **Table 5**, hypothesis testing was performed tailored to the data type of each variable. The differences in categorical variables were analyzed using Chi-square tests or Fisher's exact tests, and the continuous variables underwent two-sample t-tests to compare means between the clusters. Our analysis revealed that Cluster 1 exhibited significantly higher rates of treatment for MACE1, with 91.84% of patients receiving treatment compared to only 65.63% in Cluster 2. This difference, approaching statistical significance with p-value = 0.0597, suggests that Cluster 1 may benefit from more aggressive or effective CVD management strategies. Anthropometric measures demonstrated clear distinctions between the clusters. Hip girth measurements were significantly larger in Cluster 2 with a p-value less than 0.001, suggesting a potential correlation with increased CVD risk often associated with higher body mass. Additionally, differences in waist-hip ratio and neck girth (p = 0.0012 and p = 0.0016, respectively) further emphasize the physiological variances that might influence the cardiovascular profiles of the patients in these clusters. The clinical data highlighted substantial differences between the clusters, particularly in markers indicative of metabolic health. Creatinine and triglycerides were markedly higher in Cluster 2 (p-values < 0.0001 for both), pointing to a disturbed metabolic profile which is closely linked to increased CVD risk. These findings suggest that individuals in Cluster 2 might be at a higher metabolic risk, which could predispose them to more severe cardiovascular conditions. The analysis of demographic factors revealed a significant divergence in sex distribution and age between the clusters. Nearly all individuals in Cluster 1 were male (97.96%), compared to a lower proportion in Cluster 2 (34.38%), with a highly significant p-value of < 0.0001. Age also differed significantly, with Cluster 1 being older on average than Cluster 2 (p < 0.0001). Significant disparities were also observed in medical history, particularly in the usage of hypertension medication and the prevalence of arthritis. The use of hypertension medication was notably higher in Cluster 2 (p < 0.0001), suggesting that hypertension might be more prevalent or more aggressively treated in this group. Differences in OSA metrics were statistically significant (p = 0.0423). This highlights the critical need for effective management of sleep disorders as part of comprehensive CVD risk reduction strategies.

**Figure 5** presents a comparative analysis of the distribution distances between Cluster 1 and Cluster 2, utilizing KL divergence $D_{KL}$ to quantify the disparities in feature distributions across these clusters. Clinical data such as *"creatinine"* and *"triglycerides"* showed the most pronounced divergences, with $D_{KL} \approx 1.6$ and $D_{KL} \approx 1.2$ respectively. This suggested significant disparities in the metabolic profiles of patients in the two clusters, potentially reflecting different stages or intensities of metabolic complications related to cardiovascular health. *"Arthritis medication"* usage, a measure within the medical history category, also showed notable divergence ($D_{KL} \approx 1$), indicating substantial differences in treatment patterns for arthritis, a common comorbidity with cardiovascular diseases, between the clusters. Age exhibited a significant KL divergence at around 1.1. This suggested varying age distributions, which could influence the clusters' cardiovascular risk profiles and their response to treatment. General health scores from Zung's self-rating depression scale (both *"zung_index"* and *"zung_scored"*) showed lesser but noticeable $D_{KL}$, indicating differing levels of depressive symptoms between the clusters. Anthropometric features like *"hip girth"* also exhibited $D_{KL}$ around 0.6-0.7. Sleep-related variables such as *"mean_desat_perc"* and *"avgo2sattst"*, showed lower $D_{KL}$. However, these differences were critical in the OSA-CVD comorbidity, suggesting that while there were differences in the OSA severity, they were less pronounced than other features.

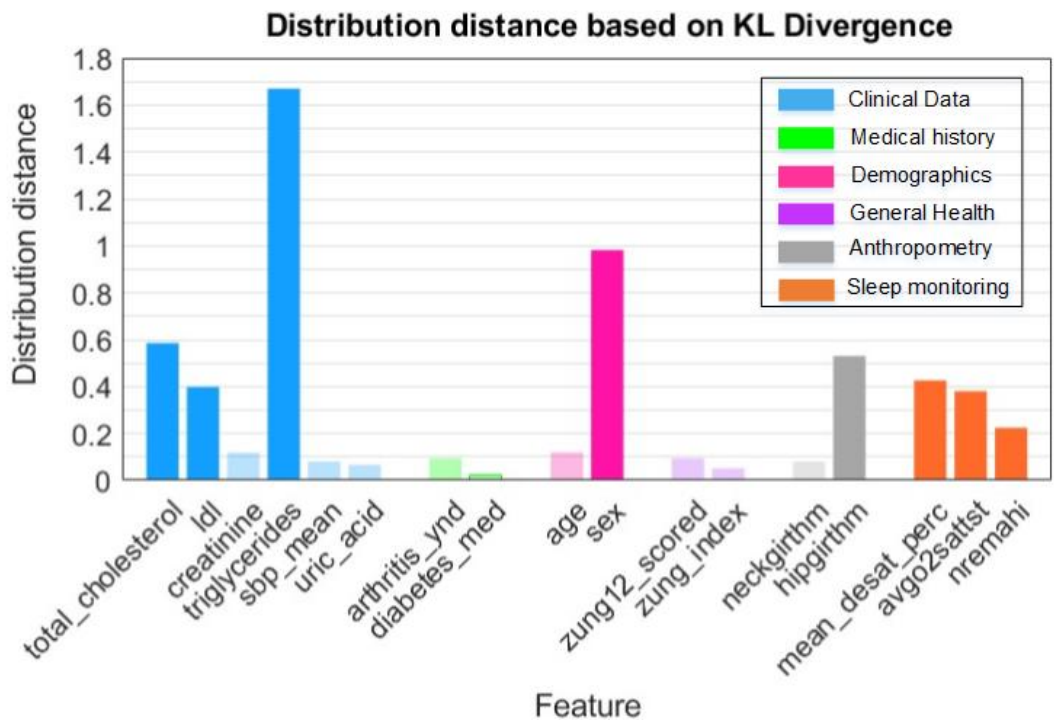

**Figure 5**. Distribution distances between Cluster 1 and Cluster 2 based on Kullback-Leibler divergence, quantifying the disparity in the distribution of key features categorized by clinical data, medical history, demographics, general health, anthropometry, and sleep monitoring.

## IV. DISCUSSION

In addressing the complex interplay of CVDs and OSA, our paper presents a comprehensive exploration of phenotype-based predictive modeling over a longitudinal timeline. The extensive data collected from the WSCS provides a robust foundation for our analysis, allowing for an intricate dissection of the progressive nature of these comorbidities through innovative statistical and machine learning techniques. Feature engineering, particularly through advanced tree-based methods, allowed us to distill the most relevant predictors from complex and high-dimensional data. This not only improved model accuracy but also enhanced our understanding of the key factors contributing to CVD risks in OSA patients. By identifying top features such as apnea-hypopnea index, levels of desaturation during sleep, and various lipid profiles, we were able to focus our analytical efforts on the most impactful variables. This targeted approach not only streamlined our modeling process but also provided clear insights into the pathophysiological links between OSA and CVD, offering potential avenues for targeted interventions. Subsequently, our approach distinguishes itself by employing a multi-level phenotypic LGMM analysis that not only identifies but also quantitatively evaluates the impact of diverse clinical, demographic, and physiological factors over time. Moreover, the utilization of the t-SNE and GMM approaches facilitates the categorization of patient data into distinct clusters, revealing significant patterns that delineate varying risk levels and progression pathways of CVDs among OSA patients. This clustering not only reflects the heterogeneity of the patient population but also enhances the specificity and applicability of the predictive models, ensuring that they are tailored to real-world clinical scenarios and can effectively inform targeted intervention strategies.

In our comprehensive analysis, we employed a methodical approach involving tree-based feature selection followed by the LGMM. The feature selection process was pivotal, isolating 20 key predictors that bridge direct and indirect relationships between OSA and CVD. The ranked features, as presented in **Figure 2**, showed that the total cholesterol, LDL cholesterol, and diabetes medications scored the highest. The fourth important feature is *"arthritis_ynd"*, indicative of a self-reported diagnosis of arthritis. However, the observed high variability in the importance score of the "arthritis_ynd" feature across different CV folds underscores the complexity and heterogeneity of the underlying inflammatory processes that are critically linked to both CVD and OSA. Studies have shown that systemic inflammation associated with chronic conditions like arthritis can exacerbate the risk and severity of OSA, further complicating the CVD outcomes in these patients [39, 51-53]. Notably, five of these features are related to sleep parameters, including the average level of oxygen desaturation during apnea/hypopnea events, average oxygen saturation during total sleep, OSA diagnosis, apnea-hypopnea index (AHI) during non-rapid eye movement (NREM) sleep, and apnea treatment. These parameters are directly linked to OSA and underscore its significance in the progression of CVD, supported by extensive clinical evidence [17, 54-56]. Furthermore, our feature selection illuminated the role of several indirect biomarkers such as depression-related scales [56], diabetes medication [57], and anthropometric variables [58], underscoring their contribution to the complex interplay between OSA and CVD. These factors, often serving as confounders, enhance our understanding of the multifaceted relationship between chronic conditions and cardiovascular risks. The comparative efficacy of the LGMM over LR in our study highlights the advantages of integrating random effects to account for temporal variations across patient visits. This is crucial as it captures the progression and fluctuation of disease markers over time, providing a more nuanced understanding of disease trajectories than traditional models, such as Cox proportional hazards models [59-61] or LR models [61].

The longitudinal LGMM analysis of our data reveals a notable enhancement in the predictive capabilities of our models across successive visits, culminating in the most robust predictions at the third visit. This trend underscores the integral role of temporal dynamics in understanding CVDs progression among OSA patients. The increasing accuracy of our predictions over time likely stems from the progressive alignment of patient risk profiles and the cumulative impact of risk factors throughout the 6-10 year follow-up period [55]. Such findings echo the patterns observed in prior longitudinal research, which consistently links OSA with elevated CVD risk. For instance, research by Wang et al. (2017) [62] highlighted the critical role of visit-to-visit blood pressure variability as a predictor for all-cause mortality, CVD incidence, and mortality, thus reinforcing the significance of longitudinal monitoring in this patient population. Our analysis further delineates two distinct phenotypes within the patient cohort, characterized by their unique CVD risk profiles. Cluster 2 poses a greater challenge for predictive modeling, attributed to its members' extensive comorbidity burden, more severe CVD manifestations, and a broader variability in risk factors. This complexity is indicative of the heterogeneous nature of CVD, which is influenced by a range of factors including metabolic conditions, lifestyle, and genetic predisposition. The clear demarcation between patients with and without CVD—based on pivotal variables like sleep apnea severity, lipid levels, diabetes status, and age—mirrors findings from Wang et al. (2006) [63]. Their study identified a constellation of factors such as prehypertension, macroalbuminuria, and obesity, which significantly amplify the risk of hypertension, a known precursor to CVD. This nuanced understanding of risk factor interplay is critical for tailoring intervention strategies that effectively address the multifaceted nature of CVD in OSA patients.

Our comprehensive analysis employing the t-SNE and GMM models has provided a profound insight into the complex trajectories and phenotypic clustering of CVD progression. The transitions observed in the t-SNE mappings across the three visits, spanning 6-10 years, were particularly telling of the dynamic nature of CVD in the context of OSA, revealing not just the evolution of individual patient conditions but also the emergence of distinct, identifiable phenotypes that evolve through the clinical timeline. Subsequent application of GMMs to these t-SNE outputs allowed for rigorous statistical modeling of these phenotypic clusters. This method effectively quantified the probability distributions of belonging to a particular cluster, thus enabling a more nuanced understanding of patient groupings based on shared characteristics and disease trajectories. Critical biomarkers such as triglycerides and hip circumference highlighted by the KL divergence, which are well-documented in their association with CVD risks [64, 65], showed significant variability across clusters. For example, our analysis linked higher triglyceride levels—associated with severe atherosclerotic conditions—to specific clusters, underscoring their role in advancing CVD pathology [66, 67]. This correlation is crucial as it aligns with emerging research suggesting that not just the presence of triglycerides, but their interaction with other metabolic factors, defines their impact on cardiovascular health. Similarly, disparities in hip circumference, which have been shown to predict cardiovascular outcomes differently based on gender, were instrumental in distinguishing between clusters, suggesting that body composition metrics can be predictive markers of CVD risk and outcomes. Moreover, the variable importance of sleep-related parameters such as the severity of hypopneas and the degree of nocturnal oxyhemoglobin desaturation was critical in delineating clusters. These findings corroborate the significant body of literature indicating the detrimental impact of disrupted sleep architecture on cardiovascular health, thereby reinforcing the integrated approaches in managing patients with OSA and CVDs [68-70].

Our study acknowledges several limitations that warrant discussion and guide future research directions. Firstly, the complexity and heterogeneity of the OSA-CVD interplay may have introduced variability in our model predictions. While we utilized advanced statistical techniques to model this complex relationship longitudinally, the intrinsic variability of individual patient responses, especially regarding intervention outcomes, might not have been fully captured. This could affect the generalizability of our findings to broader populations. Moreover, our study faced challenges with data attrition and irregular follow-up intervals, common in longitudinal studies, which could introduce bias or affect the robustness of the results. Additionally, while our models incorporated numerous relevant variables, they might not account for all possible confounders, such as socioeconomic status or environmental factors, which could influence the progression and management of CVD and OSA. Future studies should aim to incorporate more diverse cohorts that include a broader range of demographic and clinical characteristics. This would enhance the understanding of how different factors influence disease progression and treatment outcomes across various populations. Moreover, integrating novel biomarkers and more detailed genetic information could improve the predictive power of the models and offer deeper insights into the biological underpinnings of the disease processes. Furthermore, the use of real-time data collection technologies, such as wearable devices, could provide continuous, high-resolution data that better capture the dynamics of CVDs and OSA. This approach would allow for a more nuanced analysis of the interplay between sleep patterns, physiological changes, and cardiovascular health, potentially leading to more personalized and timely interventions.

## V.    CONCLUSION

In conclusion, this longitudinal study utilizing the Wisconsin Sleep Cohort has shed significant light on the intricate relationships between CVDs and OSA. Through a sophisticated multi-level phenotypic modeling approach, we have captured the dynamic interplay and progression of these comorbid conditions over an extended follow-up period. Our findings underscore the pivotal role of advanced analytical techniques like logistic mixed-effects modeling, t-SNE, and GMM in dissecting the complex interactions and trajectories of CVDs and OSA. The study highlights key predictive biomarkers and phenotypic patterns that significantly contribute to the early detection and management of CVDs in OSA patients. Notably, the identification of distinct patient clusters based on their risk profiles and disease progression pathways offers a nuanced understanding that is crucial for personalized treatment strategies. These clusters not only differentiate patients by severity and rapidity of disease progression but also align with specific therapeutic responses, thus facilitating targeted interventions. Moreover, our research emphasizes the importance of continuous, comprehensive monitoring and the integration of variable interactions over time to enhance the predictive accuracy of health outcomes in patients with OSA. This approach not only helps in identifying at-risk individuals earlier but also aids in the implementation of preventative measures tailored to individual patient profiles, potentially mitigating the progression of CVDs. Future studies should focus on expanding these findings through larger, more diverse cohorts and integrating additional predictive markers that consider the genetic, behavioral, and environmental factors contributing to the OSA-CVD comorbidity.

**DECLARATION OF COMPETING INTEREST**

The authors report no declarations of interest.